\documentclass[journal]{IEEEtran}%[journal,draftcls,onecolumn,12pt,twoside]{IEEEtranTCOM}%
\normalsize

\ifCLASSINFOpdf

\else

\fi

\usepackage{soul}
\usepackage{array}
\usepackage{color}
\usepackage{amsfonts}
\usepackage{amssymb}
\usepackage{amsmath}
\usepackage[pdftex]{graphicx}
\usepackage{cite}
\usepackage{balance}
\usepackage{caption}
\usepackage{subcaption}
\usepackage{textcomp}
\usepackage{gensymb}
\usepackage{float}
\usepackage{tabularx}
\usepackage{multirow}
\usepackage{amsthm}

\usepackage[utf8]{inputenc}
\usepackage{upgreek}
\usepackage[T1]{fontenc}
\begin{document}
\title{High Altitude Platform Station based Super Macro Base Station (HAPS-SMBS) Constellations}
\author{Md Sahabul Alam, %\IEEEmembership{Member, IEEE}, 
Gunes Karabulut Kurt, %\IEEEmembership{Member, IEEE}, 
Halim Yanikomeroglu,  
%\IEEEmembership{Member, IEEE}, 
Peiying Zhu, 
%\IEEEmembership{Member, IEEE},
and Ng\d{o}c Dũng Đào
%\thanks{}% <-this % stops a space
%\thanks{}
}
\maketitle	

%\mark{Journal of \LaTeX\ Class Files,~Vol.~14, No.~8, August~2015}

\begin{abstract}
High altitude platform station (HAPS) systems have recently attracted renewed attention. While terrestrial and satellite technologies are well-established for providing connectivity services, they face certain shortcomings and challenges, which could be addressed by complementing them with HAPS systems. In this paper, we envision a HAPS as a super macro base station, which we refer to as HAPS-SMBS, to provide connectivity in a plethora of applications. Unlike a conventional HAPS, which targets broad coverage for remote areas or disaster recovery, we envision next-generation HAPS-SMBS to have the necessary capabilities to address the high capacity, low latency, and computing requirements especially for highly populated metropolitan areas. This article focuses mainly on the potential opportunities, target use cases, and challenges that we expect to be associated with the design and implementation of the HAPS-SMBS based future wireless access architecture.
\end{abstract}
%
%\begin{IEEEkeywords}

%\end{IEEEkeywords}
%
\section*{Introduction}\label{S:Intro}
\begin{table*}
\centering
\caption{Summary of the features of a HAPS compared to a UAV and a VLEO \& LEO satellite}
\label{Table:summary}
\begin{tabular}{|l|l|l|l|l|}%{|p{6cm}|p{4cm}|p{6cm}|}
\hline 
Parameters & UAV & HAPS & VLEO & LEO\\ 
\hline \hline
 Operational altitude & $100-400$ m & $20-50$ km & $250-500$ km & $500-2000$ km\\
\hline
Cost & Low & Medium & Medium & High\\
\hline
 Round-trip propagation delay & $0.66-2.66$ $\micro$s & $0.13-0.33$ ms & $1.66-3.33$ ms & $3.33-13.33$ ms\\
\hline
Communication endurance & Short & Long & Long & Long\\
\hline
Resource limitation & High & Low (empowered by solar battery charging) & High & High\\
\hline
Mobility & Varying speeds & Quasi-stationary & Fast & Fast\\
\hline
Coverage area & Small & Wider & Wider & Wider \\
\hline
Free space path loss (dB) & $101-113$ & $147-155$ & $169-175$ & $175-187$ \\
\hline
\end{tabular}
\end{table*}
It is widely acknowledged that flexible and agile solutions for wireless connectivity will play a key role in future wireless communication systems. Currently, the connectivity requirements in terrestrial networks are addressed mainly by the densification of network infrastructures \cite{bhushan2014network}. However, densification solutions do not appear to be sufficient to address the ever-increasing user demands which are getting more and more unpredictable in space and time. In other words, no matter how dense most parts of the network is (with small base stations (BSs) in addition to macro BSs), a demanding application (such as immersed reality) can temporary arise at a locality in which the network infrastructure may be relatively sparse. In light of this, the seamless integration of terrestrial and aerial networks, known as vertical heterogeneous networks (VHetNets), has emerged as a promising architecture \cite{alzenad2019coverage}. %In general, a terrestrial network is the foundational component of an integrated network, which typically connects user equipment to the core network. 
%As the foundational component, a terrestrial network includes various network generations including 4G and 5G cellular networks.
%In the current state-of-the-art, the emergence of low earth orbit (LEO) satellite constellations have been identified as a promising solution for enhancing network coverage \cite{zhou2019coverage}. However, we note that LEO constellations have two major shortcomings: a) direct LEO to user equipment (UE) connection is difficult with the current technology due to high path loss, and b) frequent handoffs will be encountered due to the high mobility of LEO nodes. Hence, a potential complementary solution to the wireless capacity and coverage enhancement lies in aerial platforms. 
The utilization of aerial platforms for 5G wireless communication systems have already been considered in 3GPP Release 17 \cite{3gpp17}. 

The envisioned aerial network in VHetNets is composed of two interacting sub-layers which offer agile network functionalities. The first sub-layer includes moving unmanned aerial vehicle (UAV) nodes, whereas the second sub-layer is composed of  high altitude platform station (HAPS) systems, which are the quasi-stationary network elements. The International Telecommunication Union (ITU) has defined a HAPS in Article 1.66A as “A station on an object at an altitude of 20 to 50 km and at a specified, nominal, fixed point relative to the Earth". Most current deployment plans target an altitude range of 18 to 21 km. We believe that this HAPS sub-layer will provide important functions both in terms of capacity and coverage improvements by enabling the best features of both terrestrial and satellite communications. Motivated by these advantages, this paper discusses the significant role that HAPS systems can play in the future wireless access networks.
\begin{figure}[!t]
  \centering
  \includegraphics[width=\columnwidth]{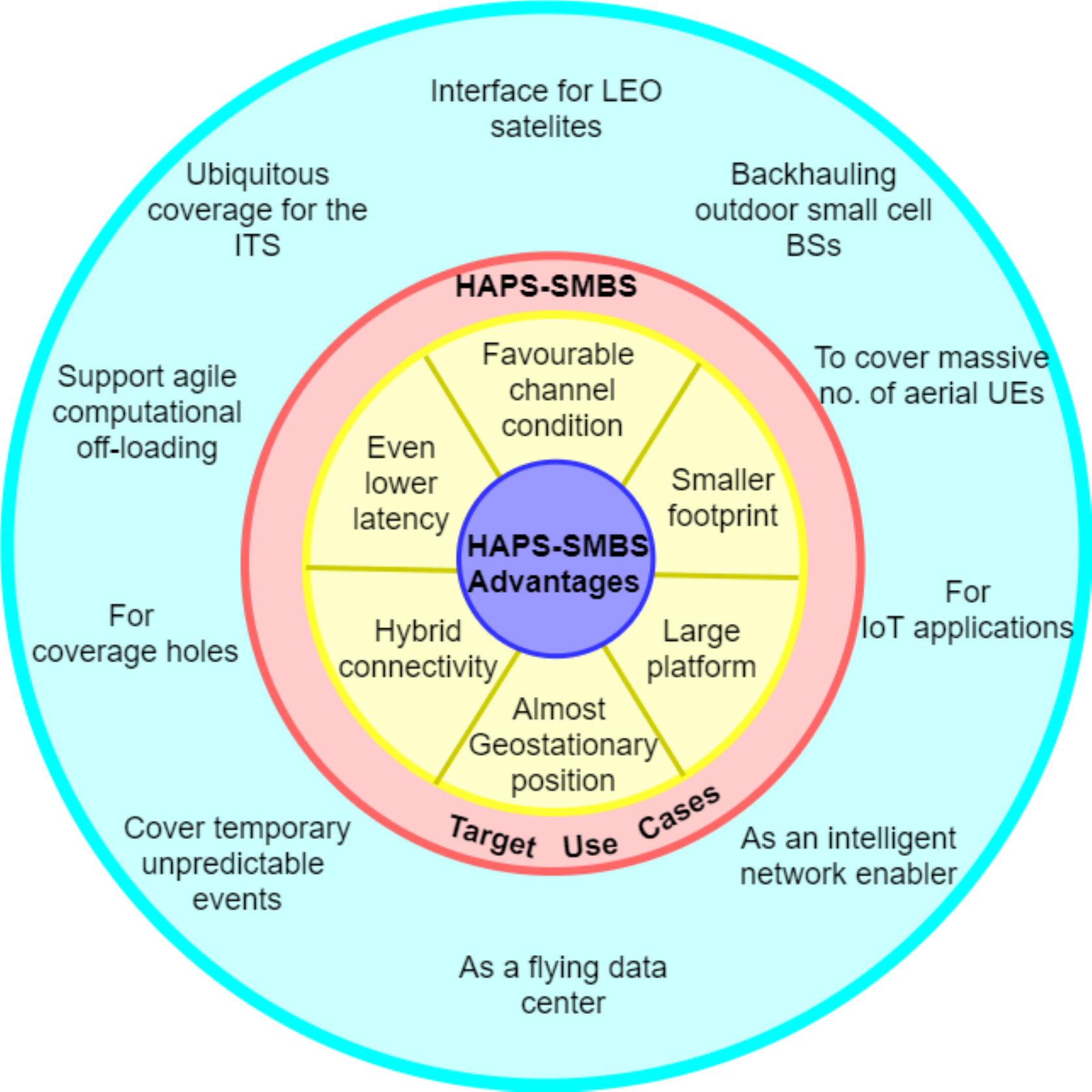}\\
  %\vspace{-10mm}
  \caption{Promises and target use cases of HAPS-SMBS networks.}\label{haps_promises}
\end{figure}

HAPS was a popular research topic in the late 1990s and early 2000s with many distinct areas of investigation \cite{arum2020review}. Despite many advantages that HAPS deployment promised, their implementation was very limited at that time. In recent years, there has been a substantial increase in research efforts and commercial application plans for different HAPS technologies \cite{arum2020review}. These developments have made HAPS systems more viable network elements owing also to the evolution of communication networks and advances in solar panel efficiency, battery energy density, lightweight composite materials, autonomous avionics, and antennas. Even supposing these advancements, all the current visions of HAPS deployments including the Google Loon project aim to bring remote parts of the globe online and for disaster applications. For example, a fresh view on the literature of HAPS from a coverage and capacity extension perspective has been given in the survey paper \cite{arum2020review}. We should note that \cite{arum2020review} mainly considers the communications issues of HAPS systems targeting remote areas. The authors in \cite{qiu2019air} considers the integration of high and low altitude platforms enabled by software defined networking in the aim to provide global access to the Internet for all in 5G and beyond, however, their proposed potential application scenarios are still limited to underserved areas. On the contrary, with the current advancements, the potential applications of a HAPS can be substantially broader than the conventional scenarios. Towards that end, this work attempts to position HAPS systems in the era of 2030s and beyond by covering various applications of HAPS for large scale communications, computation offloading, and caching. While the vision for current 2020s HAPS deployment is still to provide extended coverage for remote areas, our target use cases include densely populated areas. In this vein, the vision and framework paper \cite{kurt2020vision} intends to elaborate on the undiscovered potentials of HAPS supported by a comprehensive literature survey.
%In this vein, \cite{kurt2020vision} attempts to position HAPS systems in the era of 5G and beyond and elaborate on their undiscovered potential to serve metropolitan areas supported by a comprehensive literature survey.}

%To fulfil this research gap, this work attempts to position HAPS systems in the era of 5G and beyond by covering various applications of HAPS for large scale communications, computation offloading, and caching.}

In this article, we envision HAPS as a super macro BS, which we refer to as HAPS-SMBS, as another type of BS in a multi-tier VHetNet architecture to be deployed particularly in dense urban areas in line with the smart city paradigm. The urgency of increasing traffic volume in complex urban scenarios as well as the problems of deploying terrestrial BSs and low earth orbit (LEO) constellations motivate us to consider the deployment of HAPS-SMBS. For example, in order to provide coverage for such a metropolitan area, a large number of ground BSs, as well as a backhaul network, may be needed. This high cost of infrastructure would be a major concern compared to a HAPS-SMBS. By contrast, a HAPS-SMBS is an excellent interface to mask both high path loss and the high mobility effects of LEO constellations which have also been identified as a promising solution for enhancing network coverage \cite{zhou2019coverage}. 
%To solve the first problem, the user equipment (UE) can connect to a HAPS-SMBS with radio and a HAPS-SMBS to an LEO with free space optics (FSO). Since HAPS are almost geostationary, there are no mobility management related problems. 
The envisioned HAPS-SMBS can provide wireless services and assist the terrestrial network with the provision of distinct features, such as data acquisition, computing, caching, and processing. Fig.~\ref{haps_promises} summarizes the promises and novel target use cases of a HAPS-SMBS as a main component of wireless access architecture. This is detailed in the subsequent sections.
\section*{ Aerial Networks }\label{S:architecture}
%The aerial network is mostly connected with the terrestrial networks. 
Nowadays, aerial networks have received growing interest for their potential to improve network design both in terms of capacity and coverage. Aerial networks consist of two network components: UAV nodes and HAPS. The UAVs can be of two functions:

\textbf{UAV Base Stations:} 
%The use of  UAV-BSs is not an alternative to terrestrial BSs; they are rather a complementary solution for network management and planning. 
The coverage and capacity improvements offered through the use of UAV mounted aerial BSs is a well-studied topic in the literature \cite{alzenad20173,IREM2}. 
%This problem gained visibility in research circles with the pioneering works in \cite{alzenad20173,IREM2}. 
%Nowadays, the UAV-BS concept is more than a theoretical analysis. 
This concept is being actively investigated by 3GPP \cite{3gpp17}. The 5G system should be able to support UxNB (the 3GPP term for a UAV-BS) to provide enhanced and more flexible radio coverage. 
%On-demand capacity injection by aerial BSs for a supply-demand mismatch will be possible with these installations. Such UAVs can also act as relays to contribute towards the coverage or capacity objectives of the network \cite{IREM2}.

\textbf{UAV as User Equipment:}
The use of UAVs as user equipment (UE), such as drones %that are connected to a terrestrial network, 
is already supported through existing terrestrial networks. In particular, the use of UAVs as UE is currently being promoted by mega-retailers who would like to use the drones to carry courier packages.

%In regard to UAV communications, 
There are many survey and tutorial articles that extensively overview the UAV communications related literature. However, acknowledging distinctive features of HAPS nodes in comparison to UAVs, management of important issues in HAPS communications have very different scales compared to UAV communication systems. For example, the use of UAVs is generally time-limited, ranging from a few tens of minutes to a few hours due to onboard limited energy. In contrast, HAPS can preserve the functionality for a couple of months and preferably years. 
%On the other hand, UAVs are getting benefitted in terms of agility, flexibility, and rapid deployment in order to assist terrestrial cellular networks with reduced operational costs. Although, UAVs have these promising advantages, they are utilized for on demand short-time requirements while HAPS can be used for providing coverage and capacity in a sustainable manner. 
Moreover, while UAVs are deployed for providing communication/computation services to a small number of users on demand,HAPS can support a very large number of users in densely populated areas. Table \ref{Table:summary} summarizes the features of a HAPS compared to a UAV and a very low earth orbit (VLEO) \& LEO satellite.

To this end, an overview of how to make use of UAVs in wireless networks is out of the scope of this paper. In the following sections, we investigate the latent opportunities and challenges of HAPS in future wireless access networks. 
%To clarify, HAPS are not separable from aerial networks; nevertheless, in this paper, we focus on contributions in the field of HAPS research.

\section*{HAPS Advantages}\label{haps_advantages}
The promise of HAPS as a main component of wireless network architecture can be listed as follows.

\textbf{Favorable channel conditions:} The lower altitude combined with the high probability of line-of-sight (LoS) links provides HAPS a relatively low channel attenuation. Hence, a direct link with ground UE is possible. At a  low altitude, when compared to LEO satellites, this provides a much more favorable link budget. 
%The path loss advantage enables the use of UE as terminals which have limited transmit power levels, without the need for specialized ground stations for the uplink connectivity.
In terms of downlink, the corresponding favorable channel conditions provide a high signal-to-noise ratio (SNR) for the downlink and a capacity advantage, including for highly populated areas.
%As a HAPS is physically located between a satellite network and a terrestrial network, it can easily work as a relay, enabling satellite connectivity. This can alleviate the high gain antenna requirements for the terminals that will connect to the satellite network.

\textbf{(Almost) Geostationary positions:} The position of a HAPS is relatively stationary. This means capacity is not wasted by orbiting over unpopulated areas (e.g., oceans), at all times connectivity from the same location can be enabled. 
%Compared to a HAPS, an LEO satellite can cross over an entire continent within a matter of minutes due to their high speeds, which means that the amount of time served by that satellite is rather short. As a result, some LEO satellite communication capacity is wasted. 
The stationary status of a HAPS avoids the introduction of a  significant Doppler shift. Furthermore, no tracking of the devices are needed. The stationarity of a HAPS also provides a basis for a main mobility management node, which can contribute towards the handoff management.

\textbf{Smaller footprint compared to satellite nodes:} The smaller footprint due to lower altitudes, when compared to satellite nodes, provides a higher area throughput, and improved resource utilization capability. 

\textbf{Large platform:} A HAPS can be larger than a big building, and according to the recommendations of the ITU standard, its position should be maintained in a cylinder with a radius of $400$ m and height of $\pm 700$ m \cite{lTURF1500}. Hence it is suitable for multiple input multiple output (MIMO) and massive-MIMO (M-MIMO).
%Compounded by millimeter-wave (mmWave), spot beams 
%and highly directional antennas are possible that improve the signal-to-interference-and-noise ratio (SINR) for all users. For example, using a HAPS to cover even a temporary hot spot at the ground is possible.
%\textcolor{red}{Compounded by millimeter-wave (mmWave) with antenna array, spot beams are possible that improve the signal-to-interference-and-noise ratio (SINR) for all users. For example, using a HAPS to cover even a temporary hot spot at the ground is possible.} 
In addition, due to the large size of a HAPS, it can be equipped with wide solar panels and energy storage systems to sustain it with the energy it requires. 

\textbf{Even lower latency:} The relatively low altitude of a HAPS also provide a 40 km to 100 km round-trip distance, which corresponds to a round-trip delay of 0.13 ms to 0.33 ms. Hence, HAPS based connectivity does not suffer from the high-latency problems of satellite networks, which makes a HAPS suitable for low-latency applications.

\textbf{Hybrid connectivity:} ITU has already dedicated 600 MHz of spectrum for HAPS \cite{cianca2005integrated}. %There have been discussions about allocating an additional 3 GHz to extend the HAPS frequency band by the ITU during the World Radiocommunication Conference 2019 (WRC-19)\cite{WRC-19}. 
In addition  to the dedicated band and terrestrial cellular bands, FSO is a promising alternative for providing multi-connectivity to robust and/or high data rate communication systems. One leading solution is to generalize the multi-band radio frequency (RF) links with hybrid RF-FSO connections. 
%The FSO solutions can also serve as a backhaul through the use of dedicated ground stations for HAPS nodes. 
\section*{HAPS Super Macro Base Stations}
\begin{figure*}
\normalsize
\centering
\includegraphics[width=\textwidth]{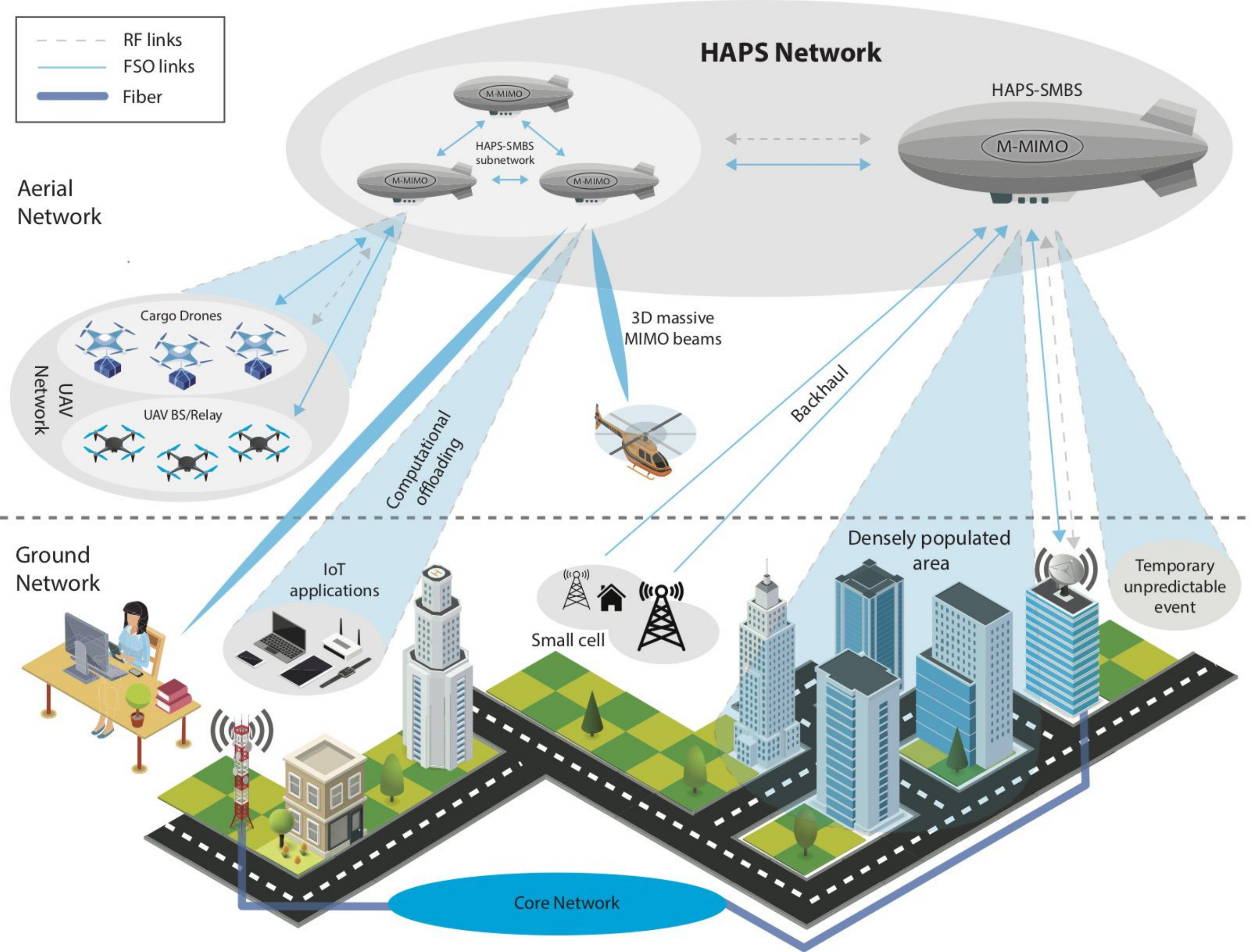}
%\vspace{-4mm}
\caption{Representation of target use cases of HAPS-SMBS networks.}
\label{HAPS_macro_BS}
\end{figure*}
A macro BS is a fundamental element in any HetNet wireless infrastructure for providing coverage and support capacity. Due to the inherent characteristics of quasi-stationarity, the smaller footprint compared to LEOs, and the LoS channels, the envisioned HAPS mounted super macro BS can serve as a powerful platform to enhance capacity, as shown in Fig.~\ref{HAPS_macro_BS}. 
%Currently, capacity requirements in terrestrial networks are addressed by a densification of the network infrastructure. However, the connection needs of mega cities are huge and growing rapidly. Hence, densification does not appear capable of addressing these ever-increasing demands. 
HAPS-SMBS improves the flexibility of the network design. The presence of HAPS-SMBS reduces the need for communication network over-engineering, which is done to match the requirements of peak demands. Therefore, the terrestrial network can be designed to satisfy the average user demands, and the rapidly changing (and often unpredictable) high demands can be simply addressed through a HAPS-SMBS. 
%Even the use of multiple coordinated HAPS-SMBS, which are equipped with multi-antenna arrays, can also enable further flexibility of the creation of separate beams in the three-dimensional space at the same time for different users through a distributed MIMO set-up. 
%Further connectivity of multiple HAPS-SMBS that cover multiple metropolitan areas can also be envisioned. 
%FSO links can also be used as the backhaul. 
It should be emphasized that we refer to this BS as a \textit{super macro BS} because of its large coverage area, enhanced capacity with M-MIMO, and the provision of supporting distinct features, such as data acquisition, computing, caching, and processing.

However, with increasing the interest in HAPS-SMBS, it is imperative to access the feasibility of its deployment mainly considering the energy consumption constraint. In this vein, there has been a successful deployment of aircraft-based solar-powered HAPS \cite{arum2020review}. The authors in \cite{arum2020energy} showed that 
%Energy management of HAPS-SMBS requires the investigation of how much energy will be consumed and how much solar energy can be harvested. For example, the authors in \cite{arum2020energy} estimated a HAPS BS power consumption for a service area radius of 60 km. It was shown that the total energy required for 24 hours of continuous HAPS mounted macro BS operation at full capacity was approximately 70 kWh. By contrast, the available solar energy provisioning using a 35 m wingspan HAPS platform is approximately 80 kWh. 
a solar power based HAPS BS for a service area radius of 60 km is potentially feasible. Additionally, significant developments in aeronautics, wingspan, and other areas of design will certainly create a HAPS-SMBS platform with higher energy efficiency.  

Future HAPS-SMBS wireless network architecture can support data acquisition, computing, caching, and processing in a plethora of application domains. Some of them are shown in Fig.~\ref{HAPS_macro_BS}, as detailed below.

\textbf{HAPS-SMBS for IoT applications:}
It is expected that in future 5G/B5G networks, HAPS-SMBS will play a key role in different applications including the internet of things (IoT). 
%In the past, there have been several research projects on HAPS; however, they are limited to civil applications, such as disaster monitoring or earth observation. 
%For example, the Google Loon project aims at delivering connectivity to people in unserved and underserved communities around the world by deploying networks through stratospheric balloons acting as flying BSs. 
IoT networks are characterized by an enormous number of devices each with low-rate links which are ideal for a single base station with a wide footprint. Due to its larger coverage, HAPS-SMBS in future 5G/B5G can support improved coverage for the realization of diverse outdoor IoT applications in a seamless, efficient, and cost-effective manner. 

\textbf{HAPS-SMBS for backhauling outdoor small cell BSs:}
Although the concept of a small cell base station has been widely acknowledged and studied for extremely high data rate coverage in 4G LTE wireless framework and is still perceived as a 5G key enabler, this concept cannot be realized in a straightforward manner mainly due to the difficulty and cost of backhauling a high number of small cell base stations. Motivated by the recent advances in HAPS and FSO research, backhauling outdoor small cell BSs can be realized through FSO and HAPS-SMBS \cite{alzenad2018fso}.
%i.e., by placing the outdoor small cell BSs wherever appropriate without much concern about backhaul, and then focusing the laser on the HAPS-SMBS for the backhaul connectivity.
\begin{figure}[!t]
  \centering
  \includegraphics[width=\columnwidth]{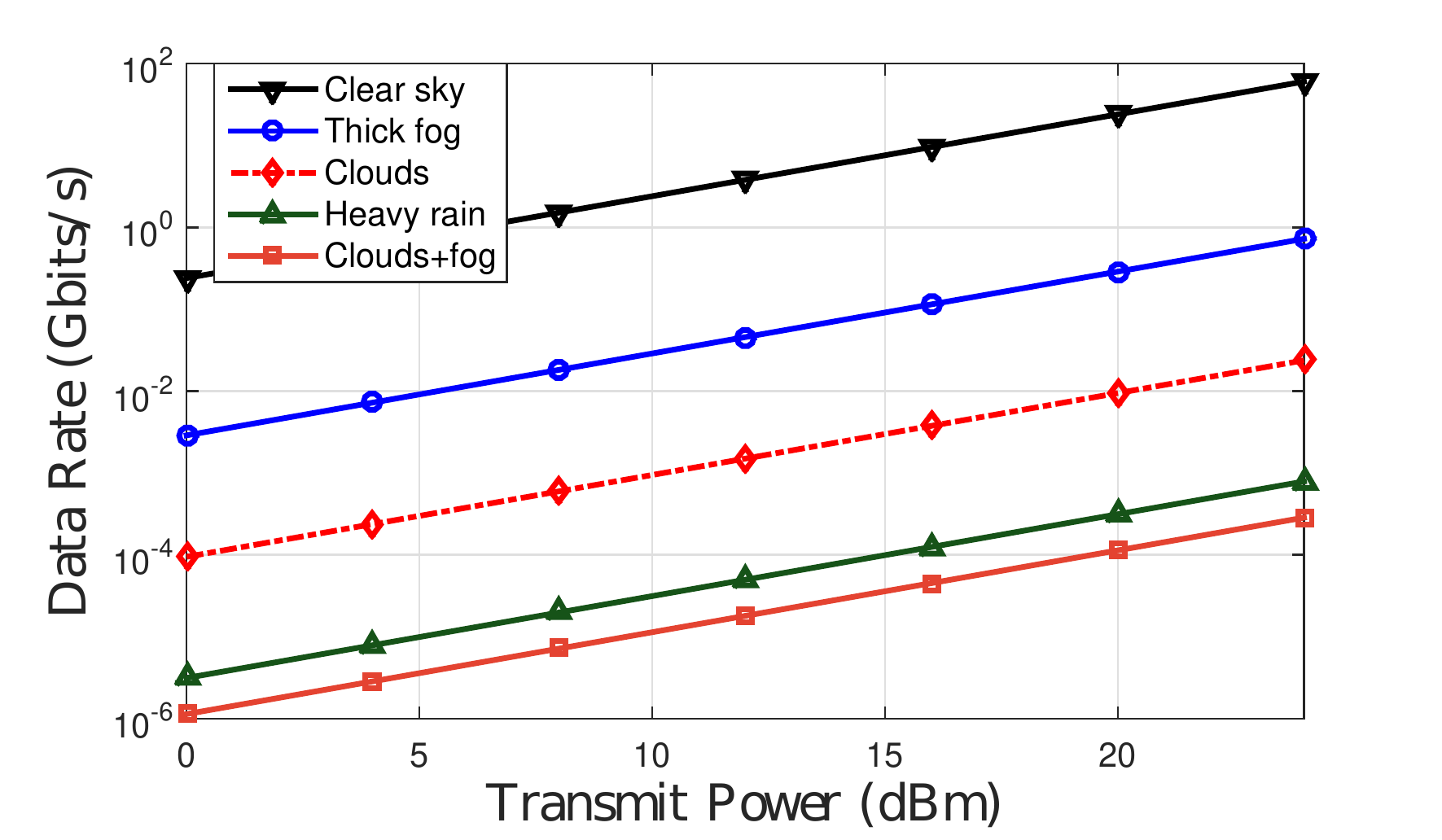}\\
  %\vspace{-10mm}
  \caption{Data rate vs transmit power of a vertical FSO link for different weather conditions. It is assumed that a HAPS-SMBS is placed at a distance of 18 km.}\label{haps_performance}
  %\vspace{5mm}
\end{figure}

For illustration, Fig.~\ref{haps_performance} shows the achievable data rate of an FSO link where it is assumed that the terrestrial small cell BSs are connected to a HAPS-SMBS through this link. The achievable data rate of a given FSO link is calculated according to \cite[Eqn. (3)]{alzenad2018fso} where the parameters listed in \cite[Table 2]{alzenad2018fso} are used to obtain the numerical results. 
%To take into account the impact of different weather conditions on the performance, we adopt the approaches developed in \cite{alzenad2018fso} for fog, rain, and cloud attenuation. 
From Fig.~\ref{haps_performance}, it can be observed that the data rates in the range of multi Gb/s can be achieved in clear weather conditions. It can also be observed that the achievable data rate is mostly affected by the rain. So the system may use FSO when there are clear skies or foggy conditions, and it can switch to RF during rainy conditions.
%\vspace{2mm}

\textbf{HAPS-SMBS to cover temporary unpredictable events:}
The proposed HAPS-SMBS based wireless network architecture can provide additional coverage in case of temporary events which are hard to predict. 
%For example, a HAPS-SMBS can provide additional beams to support the instantaneous capacity requirements in densely populated areas.
For example, a HAPS-SMBS can provide additional spot beams to support the instantaneous capacity requirements in densely populated areas. Due to high antenna gain and narrow beams, it is possible to substantially increase the data rate. Such flash events normally happen in cities, particularly when there are large gatherings, which leads to network congestion.

\textbf{HAPS-SMBS to support agile computational off-loading:}
The main idea of computational off-loading is to do the computations at the network edge near the end user in order to reduce response time and enable real-time applications. In the future, as many applications (e.g., augmented reality) will require high computational capabilities, it is expected that enabling efficient computational offloading will be a necessity. HAPS-SMBS will play a significant role in providing computational services as part of the integrated network. HAPS-SMBS have more computational power than UE (e.g., aerial UE) and can provide better coverage with LoS links due to their high position which avoids the possibility of disconnection while offloading data.
%Mobile edge computing and fog computing will play a significant role in providing computational services for end users. However, to make optimum offloading decisions, several factors need to be considered, such as the complexity of the required computations, the delay constraints, the available computational resources at the network edge, and the mobility of the end users. Therefore, several levels of computational offloading might be useful to provide different levels of computational complexity and locality (i.e., the closer to the core network the computation is done, the more computational resources can be provided with wider data scope). 

\textbf{HAPS-SMBS as a flying data center:}
HAPS-SMBS will also enable the possibility of flying data centers. These data centers can provide a back-up computational facility, that can also be functional in case of emergency scenarios where the ground infrastructure fails to function. 

\textbf{HAPS-SMBS for coverage holes:}
HAPS-SMBS can assist existing terrestrial networks by providing coverage holes through a cost-effective manner. This problem happens when the terrestrial UE received signal strength in an area falls below a predetermined level that is required for robust radio performance due to physical obstructions. For this, a HAPS-SMBS needs to steer a beam in a specific direction. 
%Through their physical advantages, HAPS-SMBS can perform 3D beamforming that enables the creation of separate beams in the three-dimensional space at the same time for different users as shown in Fig.~\ref{HAPS_macro_BS}. 
Through their physical advantages, HAPS-SMBS can perform 3D beamforming that enables the creation of separate spot beams in the three-dimensional space at the same time for different users as shown in Fig.~\ref{HAPS_macro_BS}. At present, it is widely seen that existing terrestrial networks fail to overcome the coverage holes even in a modern metropolitan area without having a viable access point to which to connect.

\begin{figure}[!t]
  \centering
  \includegraphics[width=\columnwidth]{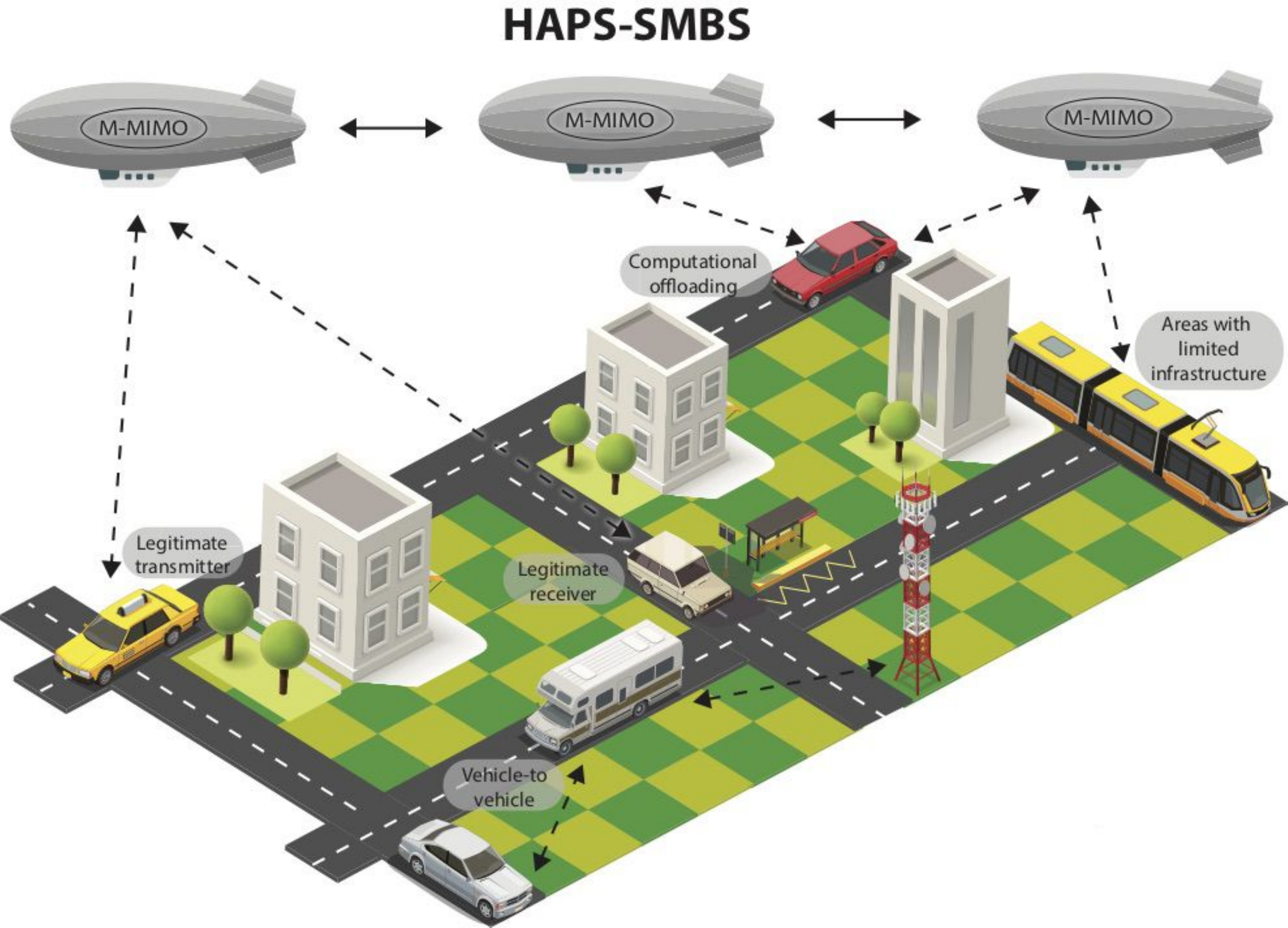}\\
  %\vspace{-10mm}
  \caption{HAPS-SMBS constellation to support ITS applications.}\label{haps_ITS}
\end{figure}

\textbf{HAPS-SMBS for ubiquitous coverage for intelligent transportation systems:}
HAPS-SMBS can play a key role for the ubiquitous coverage of intelligent transportation systems (ITS)/connected and autonomous vehicle (CAV) paradigms shown in Fig.~\ref{haps_ITS}. 
%Recent advances in sensors and the introduction of in-car wireless communication capabilities have paved the way for CAVs that enables unprecedented scenarios for road transportation. 
Nevertheless, huge data fusion and processing are necessary for many ITS applications. As vehicles are limited in processing capabilities, they may offload the data to cloud or fog computing nodes for delay-tolerant applications require large computation power. However, due to the high mobility of vehicles, data offloading will be interrupted by frequent handovers. In addition, the data processing outcome needs to be delivered through the BS, which is accessible by the vehicle. Fortunately, a HAPS-SMBS can provide both the large area coverage and computational capabilities with low communication delays. Thus, a HAPS-SMBS can eliminate the effect of frequent handovers in vehicular networks. In addition, HAPS-SMBS can be used as a sidelink for vehicle-to-vehicle communications where the terrestrial BS fails to provide enough coverage. Moreover, a HAPS-SMBS can have a wide view of a vehicular network which is essential for coordinating vehicle-to-everything communications, especially for areas with limited infrastructure.
%and optimizing the network performance.  

\begin{figure}[!t]
  \centering
  \includegraphics[width=\columnwidth]{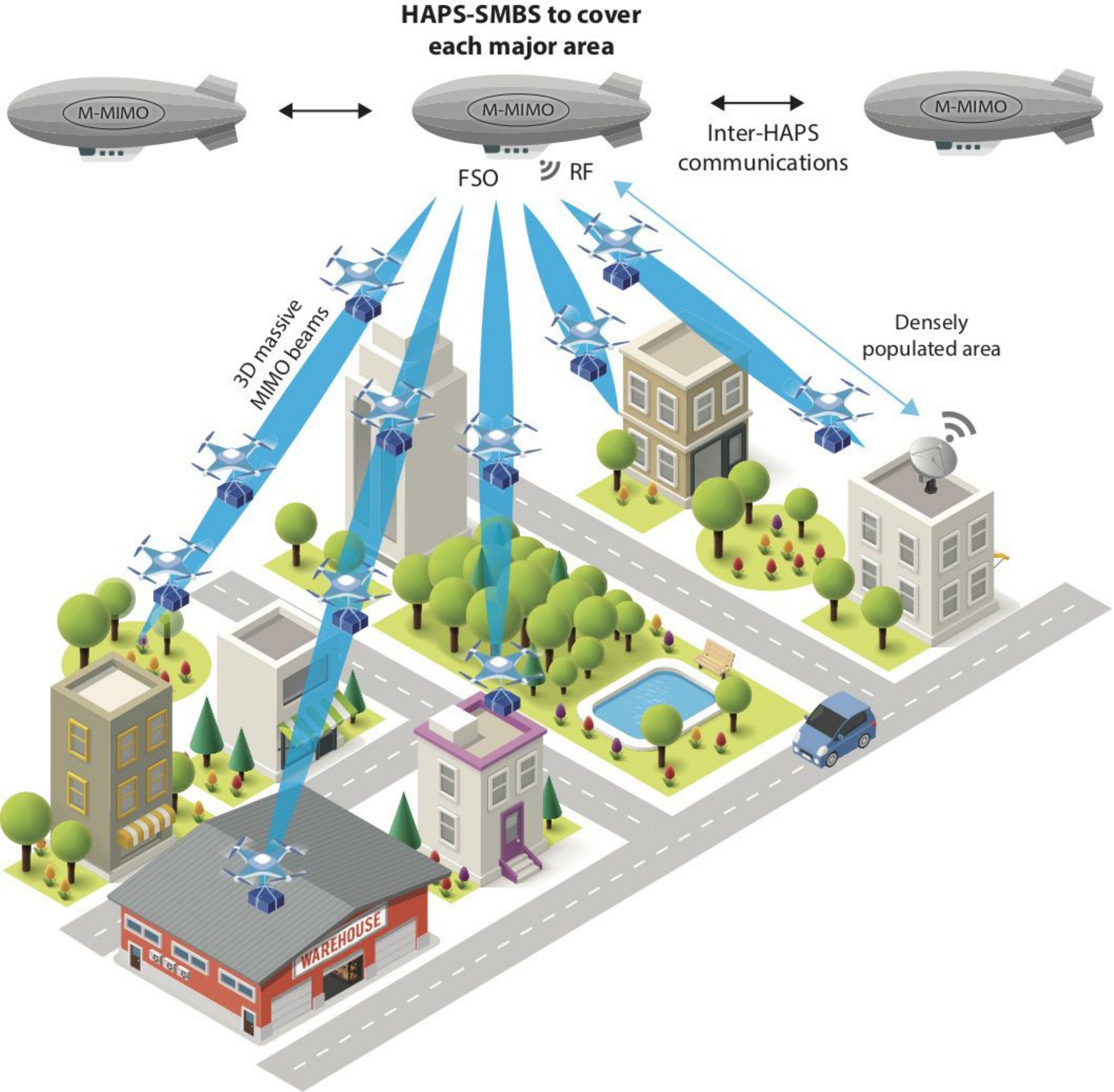}\\
  %\vspace{-10mm}
  \caption{A HAPS-SMBS for providing coverage to a massive number of aerial UE's.}\label{haps_UEs}
\end{figure}

\textbf{HAPS-SMBS to cover a massive amount of aerial UE:}
Using HAPS-SMBS for coverage can also provide an essential tool for cargo drones that will likely disrupt the retail industry in the near future. As the retail industry is evolving with the possibility of using cargo drones, 3D highways can be expected to serve the cargo package distributions of platooning autonomous drones. Considering 1 delivery/home/day for 1 million homes will require 12 drone launches per second which is the equivalent of approximately 10,000 drones in the air at any given time. To enable reliable connectivity for such a high amount of aerial UEs, a single HAPS-SMBS can be used to cover a city as shown in Fig.~\ref{haps_UEs}.

\textbf{HAPS-SMBS as an intelligent aerial network enabler:}
HAPS-SMBS can be equipped with powerful processors that can provide computational support for limited-resources aerial network elements. Besides, due to their quasi-stationary positions and large coverage areas, HAPS-SMBS can collect data from large portions of the aerial network and use such data as a real-time input for machine learning (ML) algorithms. In this way, a HAPS-SMBS can dynamically learn about the network status, resources and topology, and with a minimum dependence on terrestrial-based control, a HAPS-SMBS can control and manage the aerial network intelligently.

\textbf{HAPS-SMBS as an interface to provide seamless communication to LEO satellites:} LEO satellites move at a very high speed resulting in frequent disconnections and handovers at the terrestrial gateways. Compared to terrestrial gateways, a HAPS-SMBS has a wide upper footprint that can cover many LEO satellites simultaneously. Therefore, it is envisioned that a HAPS-SMBS can act as an interface of the satellite network and provide seamless satellite communication to the aerial and terrestrial networks. In this scenario, HAPS-SMBS will handle the frequent handover of LEO satellites, and if user devices can communicate with a HAPS-SMBS directly then users do not have to use special devices or ground stations to communicate with satellites. In this regard, supervised ML can be utilized by the HAPS-SMBS to learn the mobility patterns of the satellites in order to predict their handover then establish a connection to an approaching satellite before losing the current connection.
\section*{Open Challenges and Future Research}\label{S:Challenges}
\subsection*{Distributed RRM}
The HAPS-SMBS based wireless access architecture we envision has many distinct and even critical characteristics compared to terrestrial networks. 
%These characteristics include a high level of network heterogeneity, node mobility, frequently changed network topologies, complex channel propagation model, and SWaP (size, weight, and power) constraints. 
The distinct characteristics 
%of HAPS-SMBS based wireless architecture 
may make it inefficient to apply the standards, protocols, and design methodologies that are optimized for radio resource management (RRM) in terrestrial wireless networks in the design of HAPS-SMBS aided VHetNets directly. The RRM algorithms can be operated either in a centralized or distributed fashion. In conventional terrestrial networks, the RRM problems are typically addressed through a centralized approach. However, this may not be a feasible choice for HAPS-SMBS based VHetNets due to issues related to network heterogeneity, computational complexity, cost, spectrum overhead for channel state information (CSI) transmission, and scalability. Although each technology has distinct advantages and drawbacks, distributed RRM technology may help provide HAPS-SMBS aided VHetNets with improved agility and resilience. 

In addition, a distributed radio access network (RAN) should be coupled with distributed RRM technology to maximize its full potential. In fact, the concepts of advanced RAN and advanced RRM are inseparable. 
%For a distributed RRM, the inherent gain  relies on learning the heterogeneous environment in a dynamic and effective manner and making decisions accordingly. It is widely acknowledged that optimization approaches are the major analytical tools for RRM. 
Besides, distributed RRM should have enough cognition (cognitive radio) to decide when to transmit and which subcarriers to transmit with. Furthermore, the potential of ML should be explored in developing distributed RRM algorithms.
\subsection*{Capacity Improvement}
There are several ways to improve the capacity of communication networks. Some of them are as follows:

\textbf{Spectral efficiency improvement:} MIMO is one of the most promising techniques for improving spectrum efficiency in HAPS networks. However, many challenges have to be addressed for implementing MIMO in HAPS-SMBS. Despite the challenges, there are a considerable number of studies that have investigated the use of MIMO techniques in HAPS communications \cite{michailidis2010three}. In addition, research needs to be undertaken in aerial distributed massive MIMO, where antennas are coordinated from geographically distributed HAPS for improving spectrum efficiency. In particular, 3D MIMO, also known as full dimension MIMO, can yield higher overall system throughput.

Beamforming is also believed to have an important role in addressing the capacity demands of aerial networks at a reduced power level. Beamforming at HAPS-SMBS is more challenging than beamforming at ground BSs, where both the location and target coverage are generally fixed. Some possible directions for research on beamforming at HAPS-SMBS are the following:
\begin{itemize}
    \item 3D beamforming at HAPS-SMBS for coverage holes and unpredictable hot spots on the ground. 
    %Here, a HAPS-SMBS can cover those UEs which experience insufficient received signal power from a terrestrial BS to run a high data rate application. For this, a HAPS-SMBS needs to steer a beam in a specific direction. 3D beamforming enables the creation of separate beams in the three-dimensional space at the same time for different users. 
    \item 3D beamforming at HAPS-SMBS for aerial-UEs (such as cargo drones). %Due to the mobility of aerial UE, beamforming in such cases can be more challenging and needs to be investigated further. Also, due to the highly dynamic nature of UEs, there is a transmitter/receiver (Tx/Rx) beam alignment problem for acquiring the Tx/Rx antenna gain. Again, ML will play a great rule for getting optimal beams by leveraging side information.
\end{itemize}
Accurate beam-steering/alignment can be a challenge in moving networks; nevertheless, the quasi-stationarity of a HAPS-SMBS will help in this regard. 
%\textcolor{blue}{Nevertheless, due to quasi-stationarity of HAPS, accurate beam steering/alignment is necessary.}

\textbf{NOMA:} Non-orthogonal multiple access (NOMA) has recently been introduced as an effective approach that can potentially provide spectral efficiency, presenting a promising candidate solution for future radio systems. NOMA can also be exploited to improve spectral efficiency at HAPS-SMBS to cover a massive number of aerial UE; however, the successful operation of NOMA in HAPS-SMBS requires numerous associated challenges to be addressed, including the power coefficient determination in regard to the channel uncertainty of HAPS-SMBS to UE.

\textbf{Extension to mmWave bands:} Extending the 
spectrum to extremely high frequencies, such as mmWave bands, can be regarded as the most efficient proposal for improving transfer rates in HAPS-SMBS. In addition to the bands already dedicated for HAPS usage, for example, 47.2 – 47.5 GHz and 47.9 – 48.2 GHz, ITU during the World Radiocommunication Conference 2019 discussed that the frequency bands 21.4 - 22 GHz and  24.25 - 27.5 GHz can be used by HAPS \cite{WRC-19}. The application of mmWave techniques may offer many advantages for HAPS-SMBS, such as higher bandwidth, higher Tx/Rx antenna gain, beamforming and spatial multiplexing gain, placement of a large number of antennas in small dimensions, etc. 
%However, many challenges have to be addressed for HAPS-SMBS mmWave communication networks, including the large coverage with mmWave due \textcolor{red}{to the inherent high path loss of these bands.}
However, many challenges have to be addressed for HAPS-SMBS mmWave communication networks, including the large coverage with mmWave due to the inherent high path loss of these bands. Although, mmWave bands’ wider channel bandwidths makes the mmWave a viable solution for backhauling, nevertheless, communications via mmWave link should also be investigated for HAPS systems to UE.

%\textbf{Dynamic spectrum access:} To handle the scarcity of the radio spectrum, the concept of dynamic spectrum access deserves to be studied in HAPS-SMBS networks. Cognitive radio nodes enhance the spectrum utilization efficiency by dynamically hopping among the unused frequency bands. Moreover, due to its large coverage area and geostationary position, HAPS-SMBS can be used as a control center to hop the cognitive radio nodes optimally in a large domain.

%\subsection*{Channel Modeling}
%The underlying channel models that reflect the propagation environment will directly affect the network quality of service (QoS), and accurate channel modeling is of vital importance. Although several HAPS-ground channel models have been proposed and analyzed, further measurement campaigns for accurate channel modeling in a given environment are required to study their existence with new technologies.

%\textbf{Free space optical communication:} In addition to RF, FSO is seen as a promising and efficient technology to establish backhaul and inter-HAPS links. Through FSO technology, long-range communication links can be established. However, FSO links are very sensitive to clouds and weather conditions, turbulence, geometrical losses, and pointing losses. Therefore, accurate channel modeling for the FSO links is of vital importance to evaluate the system performance properly.

\subsection*{Network Management/Control}
The need for the joint communication, control, computing, and caching in a HAPS-SMBS to meet the intrinsic requirements of envisioned applications raises unprecedented challenges in the network management.

%In most of the existing networks, control and management approaches are centralized and will have poor performance in future networks due to their limited scalability and high response delays.
The centralized control and management approaches in most of the existing networks will likely not be suitable in future networks due to limited scalability and high response delays. There have been gradual developments to make communication networks more autonomous, self-organizing, self-configuring, and self-sustaining. To support these developments, potential solutions have been introduced in the literature; network slicing (NS), software-defined network (SDN), network function virtualization (NFV), are among them. The VHetNets architecture is highly dynamic and heterogeneous. The exploitation of NS, SDN, and NFV in the presence of a HAPS-SMBS should be explored to facilitate network reconfiguration and improve network agility and resilience. For example, the HAPS NS should consider dynamic spectrum slicing to avoid underutilization or overutilization. Furthermore, the application of machine learning techniques to derive an in-network solution in HAPS-SMBS systems is a promising research topic. The ability of having in-network solution will eliminate the need for direct human intervention on many operation levels and allow HAPS systems to make intelligent decisions in a collaborative manner.  
%There have been gradual developments to make communication networks more autonomous, self-organizing, self-configuring, and sustaining. To support these developments, potential solutions have been introduced in the literature. Network slicing (NS), software-defined network (SDN), network function virtualization (NFV), are among them. The exploitation of NS, SDN, and NFV in the integration of HAPS-SMBS in VHetNet architecture should be explored to facilitate network reconfiguration and improve network agility and resilience. Furthermore, the application of machine/deep learning techniques to derive an in-network solution in HAPS-SMBS aided VHetNet architecture would be an interesting and promising research topic.

%\textbf{Joint communication, control, computing, and caching:} Although there exist many studies on communication, network control, computing, and caching in wireless networks, in most studies these four aspects have been addressed separately. However, in HAPS-SMBS, these four "Cs" should be considered jointly to meet the intrinsic requirements of emerging applications. In this regard, the proposed HAPS-SMBS might not only be for communication purposes, but it can also provide computation and caching services and participate in the distributed network control. This scenario raises many research questions related to how the four "Cs" of HAPS-SMBS can be blended together to provide the best network performance.

\subsection*{Interference Management/Control}
HAPS were previously deployed in isolation, so there was no or few problems of interference. In metropolitan areas, one of the key challenges of deploying HAPS is interference management. In this case, owing to the simultaneous data transmission from HAPS-SMBS 
%other segments of the integrated network, more interference will be generated which may result in a higher link outage probability. 
%To access the impact of the aggregated interference produced by HAPS-SMBS on the integrated VHetNet architecture, proper interference analysis and management of interference is required. 
%In particular, for the integrated VHetNet architecture 
with terrestrial/UAV/satellite systems, more than one of these layers may be operational in the same band. For example, ITU-R allocates terrestrial cellular spectrum to HAPS systems. Hence, accurate analysis and management of the inter-layer interference due to coexistence in the same frequency band (such as mmWave bands) become crucial to guarantee the performance of the system. Furthermore, intelligent interference management is necessary, which can be achieved through implementing ML algorithms that can learn and adapt to changes in network environments. Thus, interference management through beamforming and frequency reuse can be done in an intelligent way. 
%\vspace{-4mm}
\subsection*{HAPS Constellations and Inter-HAPS Networking:} We also envision deploying a network (constellation) of interconnected HAPS-SMBSs whenever necessary. For instance, the 7,800 km long trans-Canada highway from the Pacific Ocean to the Atlantic can be served by a linear HAPS constellation towards a coast-to-coast intelligent transportation system.
%In future networks, multiple HAPS-SMBS will be deployed to cover distinct areas, and instead of working in isolation, they will form a network. 
%However, there are several challenges of applying HAPS constellation (similar to satellite mega-constellation) instead of working in isolation, for example:  
It should be noted that there are several challenges in constructing a HAPS constellation (similar to a satellite constellation) in comparison to operating a HAPS-SMBS in isolation; here are a few of those:
1) The design of a HAPS mega-constellation and its interaction with satellite mega-constellations and terrestrial network, 2) the optimization of a HAPS constellation to maximize the quality of service, and 3) the coordination among the HAPS nodes to avoid interference, wasting resources, or overlapping footprints. For instance, coordinating a HAPS network through ground stations would not be a feasible choice due to response delays, and a ground station with its limited footprint cannot have communication coverage to all the HAPS network. Therefore, we envision that HAPS networks will be self-organized with either centralised or distributed control and management system. In the centralized approach, a HAPS is elected to be the manager while the others are followers. In the distributed approach, the available HAPS in a network need to negotiate and coordinate in distributing the communication tasks. 
%in order to avoid interference, wasting resources, or overlapping footprints. 
In this regard, a comprehensive study of the HAPS constellation design methodology and inter-HAPS networking with proper interference management becomes critical.
%\vspace{-4mm}
\section*{Conclusion}\label{S:con}
In this article, we shed light on the potential opportunities and target use cases of HAPS-SMBS aided wireless access architecture. We pointed out that, while research on HAPS goes back to the late 1990s, the concept has attracted new attention in recent years, both in academia and industry, as a promising solution in future wireless networks. This momentum is fueled by the ever-increasing demand from wireless networks and also by advances in solar panel efficiency, battery energy density, lightweight composite materials, autonomous avionics, and antennas. We illustrated that the proposed VHetNets architecture empowered by HAPS-SMBS nodes will enable the network to increase the overall throughput, improve coverage, and also to provide a platform to perform the near-user computation to significantly reduce the end-to-end delays. Furthermore, HAPS-SMBS nodes will also enable the network to address unpredictable congestion instances as well as coverage holes in populated areas. 

%In a nutshell, this paper has sought to offer the reader a perspective on the novel applications of HAPS mounted BSs rather than the conventional ones.
%which is expected to become an essential component
%We showed that the proposed HAPS-SMBS with FSO transceivers will enable us to increase the overall throughput, even in densely populated metropolitan areas, to improve coverage and also to provide a platform to perform the near-user computation to significantly reduce the end-to-end delays.
%\bibliographystyle{IEEEtran}
\bibliographystyle{ieeetr}
\bibliography{maker_IEEE}

\vspace{-10mm}
\begin{IEEEbiographynophoto}{MD SAHABUL ALAM}
(sahabulalam@sce.carleton.ca)
received the Ph.D. degree in electrical engineering from ETS, Montreal, QC, Canada. Currently, he is working as a Postdoctoral fellow in systems and communications engineering department of Carleton University with prestigious FRQNT PDF fellowship. In Ph.D., Dr. Alam awarded the Governor General of Canada Gold Medal. His research interests include non-terrestrial communications and smart grid communications.
\end{IEEEbiographynophoto}
\vspace{-12mm}
\begin{IEEEbiographynophoto}{GUNES KARABULUT KURT}
(gkurt@itu.edu.tr) received the Ph.D. degree in electrical engineering from the University of Ottawa, Ottawa, ON, Canada, in 2006. 
%Between 2005 and 2008, she was with TenXc Wireless, and Edgewater Computer Systems, in Ottawa Canada. From 2008 to 2010, she was with Turkcell R\&D Applied Research and Technology, Istanbul. 
Since 2010, she has been with ITU. She is also an Adjunct Research Professor at Carleton University. She is serving as an Associate Technical Editor of IEEE Communications Magazine.
\end{IEEEbiographynophoto}
\vspace{-12mm}
\begin{IEEEbiographynophoto}{HALIM YANIKOMEROGLU}
[F] (halim@sce.carleton.ca) is a full professor in the Department of Systems and Computer Engineering at Carleton University, Ottawa, Canada. His research interests cover many aspects of 5G/5G+ wireless networks. His collaborative research with industry has resulted in 38 granted patents. He is a Fellow of the Engineering Institute of Canada and he is a Distinguished Speaker for IEEE Communications Society and IEEE Vehicular Technology Society.
\end{IEEEbiographynophoto}
\vspace{-12mm}
\begin{IEEEbiographynophoto}{PEIYING ZHU}
[F] (peiying.zhu@huawei.com) is a Huawei Fellow. She is  currently leading 5G wireless system research in Huawei. The focus of her research is advanced wireless access technologies with more than  150 granted patents. In recent years she organized and chaired various  5G workshops. She is a frequent keynote speaker in major IEEE conferences.%DũNG Đào
\end{IEEEbiographynophoto}
\vspace{-12mm}
\begin{IEEEbiographynophoto}{NG\d{O}C D\~UNG Đ\'AO}
(ngoc.dao@huawei.com) is a principle engineer of Huawei  Technologies Canada. His research interest covers several aspects of 5G and beyond 5G mobile networks, including architecture design, data  analytics, and vertical applications. He is a technical editor of IEEE  Communications Magazine, associate editor of IEEE Communications Surveys \& Tutorials, and editor of IEEE Transactions on Vehicular Technologies.
\end{IEEEbiographynophoto}

\end{document}